\documentclass[aps,prb,twocolumn,showpacs,groupedaddress]{revtex4-2}

\usepackage{graphicx}

\begin{document}

\title{Coordinate and Momentum Distributions of a Small Composite System}
\author{$^{1}$\textbf{Dmitry Naplekov},$^{1}$\textbf{Yury Bezruk},$^{1,2}$\textbf{Vladimir Yanovsky}}
\affiliation{$^1$ Institute for Single Crystals, NAS Ukraine, 60 Nauky Ave., Kharkov, 61072, Ukraine}
\affiliation{$^2$ V. N. Karazin Kharkiv National University, 4 Svobody Sq., Kharkiv, 61022, Ukraine}

\begin{abstract}
An isolated small system consisting of a movable shell and a finite number of internal particles is considered. Exact distributions of the shell coordinates and momenta are obtained for various types of internal particle motion under the conditions of conservation of the total energy and momentum of the system. The coordinate distribution consists of a finite number of branches and may contain plateau regions. The momentum distribution corresponds to a non-uniform distribution of energy among the degrees of freedom of the system. It is shown that the coordinate distribution depends on the number of internal particles and remains unchanged when the dimensionality of their motion varies, whereas the momentum distribution depends on the number of degrees of freedom. Consequently, comparison of the coordinate and momentum distributions of the shell makes it possible to draw conclusions about the internal structure of the system. The chaotic motion of the shell caused by impacts from the internal particles is analyzed; unlike Brownian motion, it persists even in the absence of an external medium. A universal expression for the root-mean-square deviation of the shell is obtained explicitly, depending on the number of internal particles and their masses.
\end{abstract}
\pacs{05.45.-a}
\bigskip

\maketitle

\section{Introduction}

Interest in small systems with a finite number of degrees of freedom arises both from the existence of numerous systems belonging to this class and from the unusual properties they exhibit. Such systems include traditional macromolecules and microclusters. This list has been significantly expanded by nano-objects, among which compounds of a new type --- rotaxanes \cite{ODO-Sch} --- should be mentioned. These artificial chemical compounds with topological bonding were synthesized relatively recently \cite{ODO-Har}. They represent mechanically interlocked supramolecular systems in which ring components are threaded onto linear axial components. Even more complex objects are formed by nested nanotubes and fullerenes. As an example, one may mention nanopeapods  \cite{ODO-Smi,ODO-Mon,ODO-Yud}. Such a small system consists of fullerene $C60$ molecules placed inside a single-walled carbon nanotube. The dynamics of confined systems represents a serious problem both for fundamental physics and for applied research and has been investigated experimentally \cite{ODO-Rol,ODO-Kit,ODO-She}. The general difficulty is that such systems cannot be described using the thermodynamic limit.

As noted above, small systems exhibit properties associated with finite size and a finite number of degrees of freedom. These properties have been observed experimentally. For example, the decrease in the melting temperature of nanoclusters with decreasing cluster size is well established  \cite{ODO-Wro,ODO-Ber,ODO-Buf,ODO-Lai,ODO-Zha}, as is the coexistence of solid and liquid phases in nanoparticles. This unique state (a hybrid phase) arises because, due to the small size and the strong influence of surface energy, melting occurs gradually rather than instantaneously. Nanoparticles may exist in a quasi-liquid state, possessing a solid core and a liquid shell \cite{ODO-Cou,ODO-Los,ODO-Khsh}. Predictions that the microcanonical heat capacity of a small system can become negative have also been confirmed experimentally. In particular, it has been demonstrated that a cluster containing exactly 147 sodium atoms indeed possesses a negative microcanonical heat capacity in the vicinity of the transition from the solid to the liquid state  \cite{ODO-Schm}. Of course, other unusual properties of small clusters have also been discovered, including magnetic effects and others.

Quite often, systems containing a finite number of particles exhibit atypical features in their momentum distributions. In a number of cases such distributions arise when the principle of maximal non-additive entropies is employed  \cite{ODO-Tsa,ODO-Tsal,ODO-Saa,ODO-Pla}. Such systems also include certain Hamiltonian systems for which a microscopic approach has been developed  \cite{ODO-Ray,ODO-Rom}. In \cite{ODO-Ray} it was shown that the momentum distribution in small systems described by the microcanonical ensemble is not Maxwellian. In a certain sense, this work develops earlier ideas of Maxwell \cite{ODO-Max}. In the system considered here, similar anomalies may be expected, although the presence of a shell and the specific mechanism of energy exchange with it significantly influence the resulting distributions.

Although all physical systems are, in principle, in contact with the surrounding environment, small systems are often located in vacuum, and on the time scales of interest they may be considered isolated and in equilibrium. In such cases, besides the total energy, additional conserved quantities may exist. Therefore, considerable attention is currently devoted to studying the general properties of small systems  \cite{ODO-Nii}, their statistical distributions  \cite{ODO-Sca,ODO-Liu}, and the influence of additional integrals of motion on these distributions \cite{ODO-Nii1}. The appearance of additional integrals of motion in molecular-dynamics simulations is often associated with the use of periodic boundary conditions. The effect of such additional integrals of motion can be significant when the number of degrees of freedom is small \cite{ODO-Xu}, gradually diminishing as this number increases.

It is common for studies of small systems to focus on the distributions of their energies and momenta. In the present work we also investigate the distribution of their coordinates. Such distributions are related to how well the position and shape of the system boundary are defined. This is important, for example, when separating the density profile of the system boundary from the motion of the boundary itself  \cite{ODO-Hie,ODO-Her,ODO-Fuk}, including in the case of nanodroplets in connection with the problem of how the surface tension coefficient changes as the droplet size decreases. Coordinate distributions and finite-degree-of-freedom effects are also relevant, for example, to scattering cross sections in heavy-ion collisions  \cite{ODO-Kuz,ODO-Sor}.

In this work, analytical expressions are obtained for the coordinate and momentum distributions of a structurally complex particle consisting of a shell of a given mass and a finite number of internal particles with different masses. It is shown that the coordinate distribution has a complicated structure and depends strongly on the ratios of the masses of the internal particles. The existence of a coordinate distribution function makes it possible to establish the properties of the chaotic motion of the shell. This chaotic motion is caused by impacts from the internal particles and, unlike Brownian motion, persists even in the absence of a surrounding medium. A universal expression for the standard deviation of the shell is obtained explicitly; it depends on the number of internal particles. A comparison of simulation data with the analytical results obtained demonstrates good agreement. The influence of additional conservation laws on the form of the distribution functions is discussed. It is shown that the momentum distribution of the shell is determined by the shell mass and the total mass of the internal particles. In this case, the equipartition of energy among the degrees of freedom is violated. The specific features of the resulting distribution functions are analyzed and discussed.

\section{The System Under Consideration and Its Integrals of Motion}

\begin{figure}
 \centering
 \includegraphics[width=7 cm]{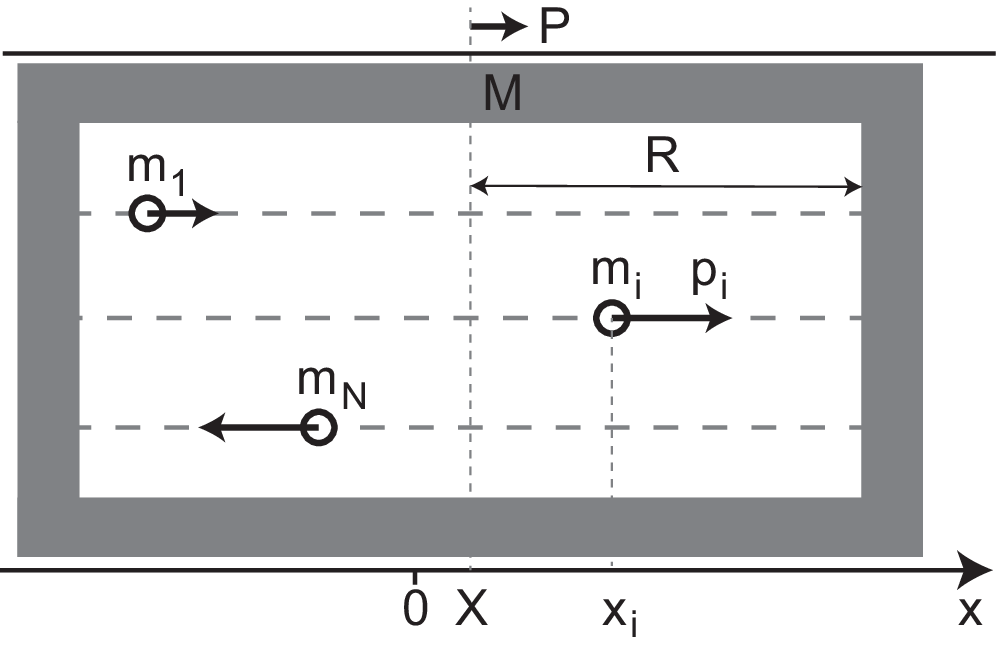}
 \includegraphics[width=7 cm]{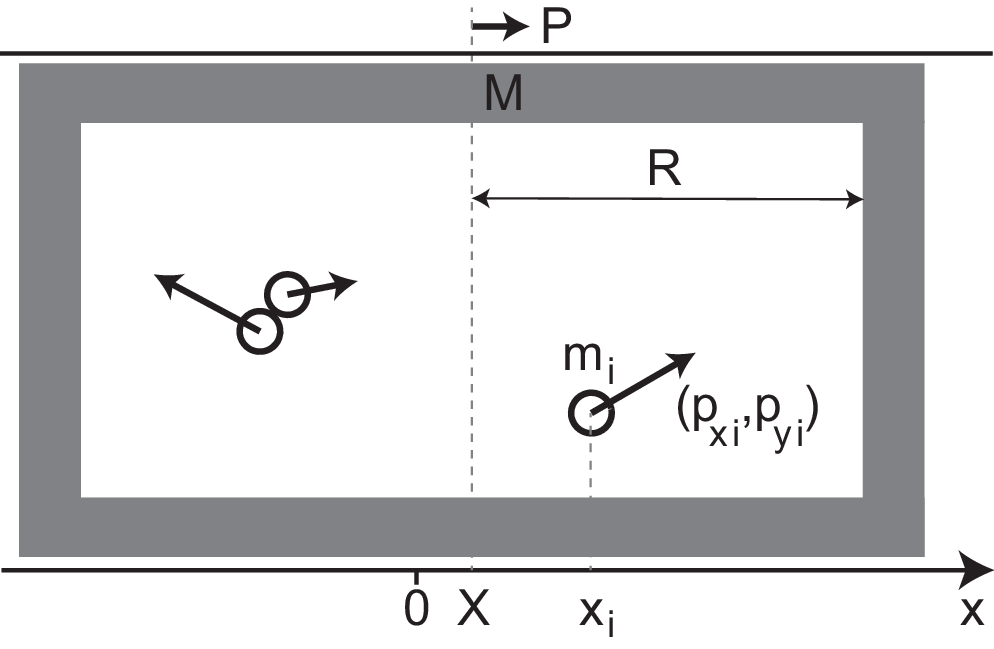}\\
 \caption{A movable shell with mass of $M$ and length of $2R$, containing a finite number  $N$ of particles with masses $m_i$. The shell performs one-dimensional motion along the  $x$-axis. Both one-dimensional and two-dimensional motions for the internal particles are considered. The total energy of the system $E_{tot}$ is finite, the total $x$-component of the momentum $P_{tot}$ is equal to zero, and the position of the center of mass $X_{tot}$ coincides with the origin of the coordinate system.}
 \label{ODO-fig1}
\end{figure}
We consider an ideal gas consisting of a finite number $N$ of particles confined within a movable shell (Fig.\ref{ODO-fig1}). The shell can move one-dimensionally along the $x$  axis, has mass of $M$, momentum of $P$ and size of $2R$. For the internal particles we analyze two possible types of motion:one-dimensional motion without particle-particle collisions and two-dimensional motion with interparticle collisions (Fig.\ref{ODO-fig1}). Let the particle coordinates be $x_i$ (and in the two-dimensional case also $y_i$), their momenta $p_{x i}$ and $p_{y i}$, and radii $r_i$. The particle masses $m_i$ may in general differ; their total mass is denoted by $m_{tot}=\sum_{i=1}^{N} m_i$. All collisions between particles and with the shell are assumed to be perfectly elastic. The first scenario can be regarded as a special case of the second corresponding to particular initial conditions.

The system always possesses at least two integrals of motion. During its evolution the following quantities are conserved:total energy, total momentum and, as a result, center-of-mass position. We assume the total momentum  $P_{tot}=0$, the center of mass $X_{tot}$ located at the origin, and the total energy equal to $E_{tot}$. This corresponds to describing a structurally complex particle in the center-of-mass reference frame. In the case of colliding internal particles, no additional integrals of motion exist because collisions induce a high level of dynamical chaos. If particles interact only with the shell, the degree of chaoticity is lower and additional integrals of motion may arise. As will be shown below, when $N=2$ the system possesses one more integral of motion, which disappears for larger numbers of particles.There also exists a degenerate case when all particle masses are equal and equal to the shell mass. In this situation each collision simply exchanges momenta between the particle and the shell. For any number of particles, the momentum distribution of the shell or particles becomes discrete and corresponds to the initial set of momenta.

A statistical description of the system must take into account all integrals of motion, since each affects the form of the statistical distributions. It is also necessary to enforce the constraint that particles remain inside the shell, i.e. $X-R<x_i<X+R$. Coordinate distributions are influenced only by conservation laws involving coordinates-primarily the conservation of the center-of-mass position. Conversely, momentum distributions depend on the conservation of energy and momentum. In what follows we derive these distributions for the shell.

\section{General Statistical Description}

To obtain the distribution of a particular phase-space variable, one must first determine the full microscopic distribution and then integrate it over all variables except the one of interest. For two conserved quantities, the microcanonical distribution is typically written as:
\begin{equation}\label{ODO-1-1}
\rho=const \; \delta (E_{tot}-E(p,q)) \, \delta (P_{tot}-P(p,q))
\end{equation}
To transform this expression into ordinary functions, we divide the phase variables into independent ones and two dependent ones, chosen as the momentum components $p_i$ and $p_j$. For independent phase variables, one can obtain a microscopic distribution in terms of ordinary (rather than generalized) functions, while the remaining two components of the momenta are determined by all the other variables and by the conservation laws.

A Laplace transform of Eq. \ref{ODO-1-1}, with respect to  $E_{tot}$ and $P_{tot}$ yields

\begin{equation}\label{ODO-1-2}
\pounds \rho=const \; e^{-E(p,q) s_1 - P(p,q) s_2} \: \Theta (E(p,q)) \Theta (P(p,q))
\end{equation}

The energy of the system is a priori a positive quantity; therefore, its $\Theta$-function is equal to 1. The momentum of the system, without loss of generality, will also be taken to be nonnegative.

To obtain the distribution of the independent phase variables, this expression must be integrated over the two eliminated phase variables $p_i$ and $p_j$, followed by an inverse Laplace transform. Let us single out these variables from the conservation laws for the energy and for a component of the momentum as follows:

\begin{equation}\label{ODO-1-3}
\begin{array}{l}
E(p,q)=\frac{p_i^2}{2 m_i}+\frac{p_j^2}{2 m_j}+E_1\\
P(p,q)=p_i +p_j +P_1\\
\end{array}
\end{equation}

Then the integral over the variables $p_i$ and $p_j$ will look like this:
\begin{equation}\label{ODO-1-4}
\pounds \rho_{proj}=C \int\limits_{-\infty}^{\infty} \int\limits_{-\infty}^{\infty} e^{-(\frac{p_i^2}{2 m_i}+\frac{p_j^2}{2 m_j}+E_1) s_1 - (p_i +p_j +P_1) s_2} dp_i dp_j = 
\end{equation}
	\[=\frac{C_1}{s_1} e^{-E_1 s_1 - P_1 s_2+(m_i +m_j) \frac{s_2^2}{2 s_1}}
\]
After performing the inverse Laplace transform over $s_2$:

\begin{equation}\label{ODO-1-5}
\pounds \rho_{proj}=const \frac{e^{- s_1 (E_1+\frac{(P_{tot}-P_1)^2}{2(m_i +m_j)})}}{\sqrt{s_1}}
\end{equation}

After the inverse Laplace transform over $s_1$ we obtain the distribution:

\begin{equation}\label{ODO-1-6}
\rho_{proj}= \frac{const}{\sqrt{2 (E_{tot}-E_1)(m_i +m_j )-(P_{tot}-P_1)^2}}
\end{equation}

This expression represents the probability density for the particle coordinates $x_1 \ldots y_N$ and momenta $p_{x 1} \ldots p_{y N}$ excluding two eliminated momentum components $p_i$ and $p_j$. In what follows, we will choose $p_{x N-1}$ and $p_{x N}$as these components. Eq.\ref{ODO-1-6} can also be obtained without using the microcanonical distribution Eq.\ref{ODO-1-1}, directly as a consequence of Liouville's theorem for Hamiltonian systems of particles. Such a derivation is analogous to the one we employed in our previous works \cite{ODO-my,ODO-my1} for the case of conservation of energy and angular momentum of the system.

\section{Distribution of the Shell Coordinates}

Our primary interest is the behavior of the shell as a collective variable. Therefore, we consider the distribution of the shell coordinate  $\rho(X)$. To obtain this distribution, Eq.\ref{ODO-1-6} must be integrated over the accessible region of phase space. This region of phase space is finite, since for a finite total energy the law of conservation of energy imposes bounds on the maximum possible momenta of the particles and the shell. The conservation of the center of mass and the size of the shell constrain the maximum possible coordinates of the particles and the shell. Thus, all phase variables vary within finite limits. We assume that a typical trajectory uniformly fills the accessible phase-space region. The validity of this assumption will be verified by comparison with numerical simulations.

Because coordinates do not appear explicitly in the conservation laws (except through the center-of-mass condition), the microscopic probability density does not depend on them. For the system under consideration, this property also holds. Therefore, the analytical form of the coordinate distributions-particularly that of the system's shell-can be obtained in a relatively straightforward manner. Consequently, after integrating Eq.\ref{ODO-1-6} over all momenta, the probability density becomes constant. That is, the probability of finding the system in a state characterized by given coordinates of the particles and the shell does not depend on the specific values of those coordinates. The coordinate distribution of the shell $\rho(X)$ is therefore obtained by integrating this constant over all admissible positions of the internal particles. Since the coordinate of one of the particles is not independent, the number of integration variables is one less than the number of internal particles. This reflects the influence of the conservation of the center-of-mass position, which also manifests itself in the form of the integration limits. The coordinate of the last particle is uniquely determined by this conservation law. This leads to:

\begin{equation}\label{eq1}
\rho(X)=\int_{x_{1 min}}^{x_{1 max}} .. \int_{x_{N-1 min}}^{x_{N-1 max}} const \quad dx_1 .. dx_{N-1}
\end{equation}

It is worth noting that this distribution applies to both cases of particle motion inside the shell considered here. If motion of the shell along other axes were possible, this would not affect the distribution of its $x$-coordinate.  This distribution is also unaffected by either a change in the dimensionality of the internal particles, motion or by the value of their average energy.

The distribution \ref{eq1} is non-zero within the interval  $X \in [-\frac{\sum_{j=1}^{N} m_j}{M+\sum_{j=1}^{N} m_j} R ,\frac{ \sum_{j=1}^{N} m_j}{M+\sum_{j=1}^{N} m_j}R] $, where the maximum allowable displacement is determined by the center-of-mass conservation law. For shell coordinates outside this range, the distribution is equal to zeroThe limits of integration for the internal particles are determined by two conditions: the positions of some particles must allow the correct position of the center of mass to be established by the remaining particles, and all particles must remain inside the shell. Taking these conditions into account, the limiting permissible values of the coordinate of particle  $x_i$ for given coordinates of particles from $x_1$ to $x_{i-1}$, can be readily obtained as:

\begin{equation}\label{eq2}
\begin{array}{l}
x_{i min}=Max(X-R, -\frac{M X+(X+R)\sum_{j=i+1}^{N} m_j +\sum_{j=1}^{i-1} m_j x_j}{m_i})\\
x_{i max}=Min(X+R,-\frac{M X+(X-R)\sum_{j=i+1}^{N} m_j +\sum_{j=1}^{i-1} m_j x_j}{m_i})\\
\end{array}
\end{equation}
\begin{figure}
 \centering
 \includegraphics[width=6 cm]{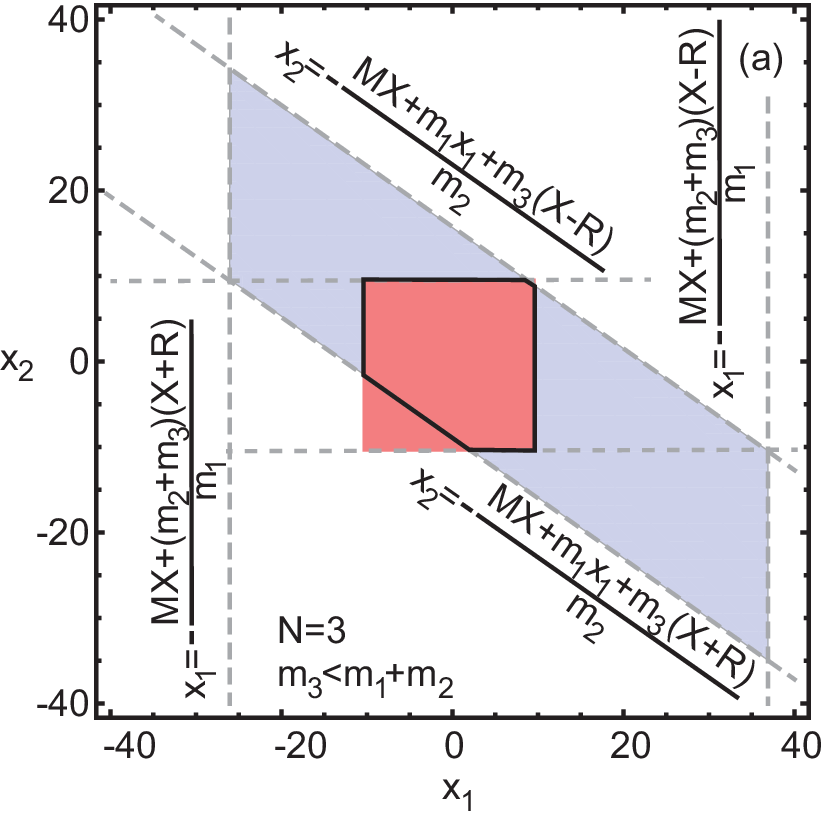}
 \includegraphics[width=6 cm]{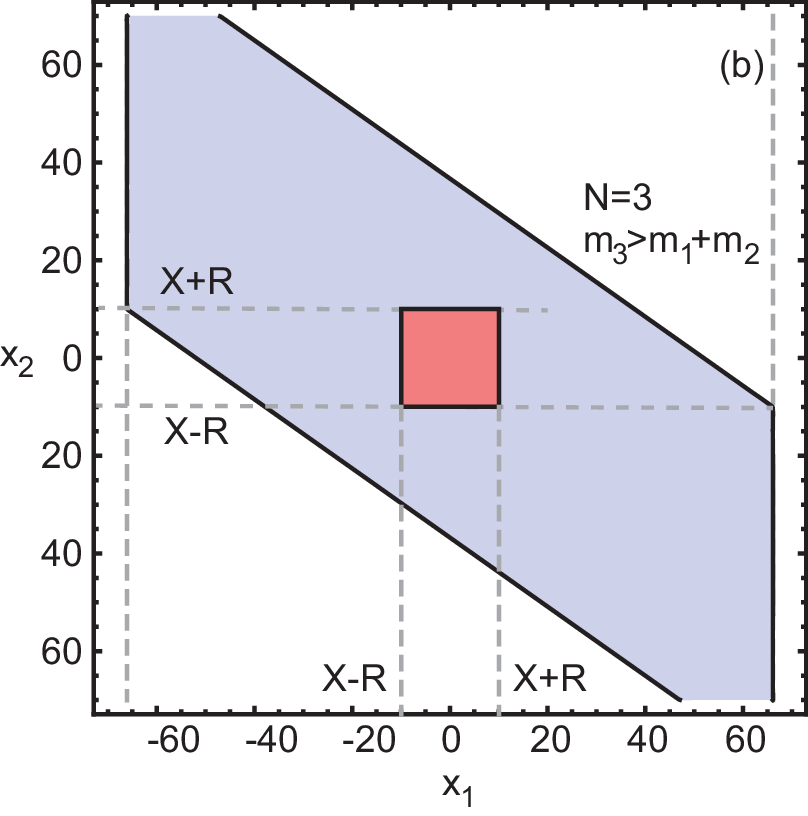}\\
 \includegraphics[width=6 cm]{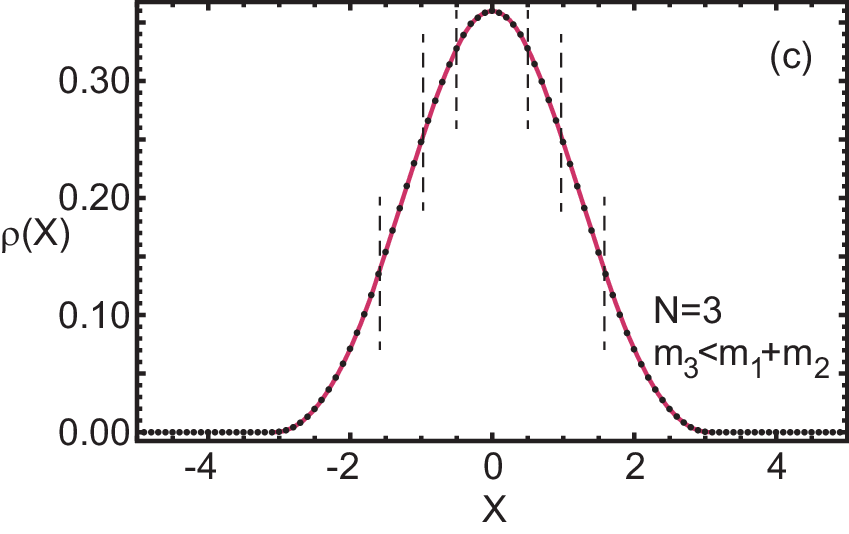}
 \includegraphics[width=6 cm]{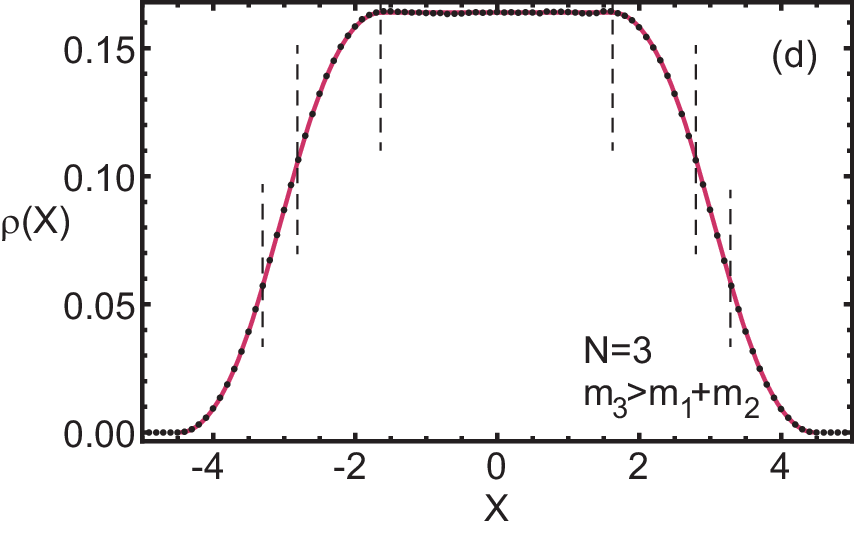}\\
  \caption{(a,b) Integration domains in the case of  $N=3$ internal particles, depending on whether the condition $m_3 > m_1+m_2$is satisfied. As  $X$ increases, the square moves to the right while remaining between the horizontal lines $x_2=X \pm R$, passing through the intersection points of the strip boundaries. (c,d) Distributions of the shell coordinate corresponding to the area of intersection of the square and the strips. The black dots represent the results of numerical simulations, while the red solid line shows the distribution given by Eq. \ref{eq3}. The vertical dashed lines indicate the boundaries between the segments of the distribution. System masses:  $M=3 \pi$, $m_1=1$, $m_2=\sqrt{2}$, (a,c) $m_3=\sqrt{3}$ and (b,d) $m_3=3\sqrt{3}$, shell dimensions $R=10$, $H=10$.}
 \label{ODO-fig2}
\end{figure}

These integration limits contain a conditional constraint; therefore, for some values of  $X$ the limits have one analytical form, whereas for other values they have a different form. As a consequence, the distribution of the shell coordinate consists of several distinct segments. For a sufficiently large number of particles, the form of this distribution is most conveniently obtained numerically using Eqs. \ref{eq1} and \ref{eq2}. If the number of internal particles is small, it is also possible to derive an explicit analytical form for all segments of the distribution and their boundaries. Let us consider in more detail, for example, the distribution of the shell coordinate in the case of three internal particles. Without loss of generality, the particle masses may be assumed to be ordered as  $m_1 \leq m_2 \leq m_3$. Since the integration is performed over a constant with respect to two variables, the value of the distribution equals the area of the integration domain For a fixed shell coordinate  $X$, the coordinates of all particles must lie within the  $X-R<x_i<X+R$ interval. In addition, the coordinate of the first particle cannot deviate from the center of mass in the positive direction by more than $x^{\{ 1 \}}_{Max}$, which is obtained from  $M X+m_1 x^{\{ 1 \}}_{Max}+(m_2+m_3)(X-R)=0$. An analogous expression gives the maximal permissible deviation of the coordinate in the negative direction. For a fixed position of the first particle, the coordinate of the second particle cannot deviate from the center of mass in the positive direction by more than  $x^{\{ 2 \}}_{Max}$, which is obtained from $M X+m_1 x_1+m_2 x^{\{ 2 \}}_{Max}+m_3(X-R)=0$. The maximal negative deviation is obtained in a similar manner. As a result, for a fixed shell coordinate $X$ the integration domain represents the intersection of two strips and a square, as shown in Fig.\ref{ODO-fig2}(a,c). As $X$ varies, the positions of these strips and the square change; however, the square always remains between the horizontal lines $x_2=X \pm R$, passing through the intersection points of the boundaries of the two strips. The unnormalized value of the distribution $\rho^{\{ N=3 \}}(X)$ is equal to the area of the intersection of these three regions.

To write the analytical form of the shell coordinate distribution for three internal particles, we first note that the square may either not be entirely contained within the intersection of the two strips (see Fig.\ref{ODO-fig2}(a)), or may lie completely inside this intersection (see Fig. 2(c)). The first situation corresponds to the condition  $m_3 < m_1+m_2$, and the second to $m_3 > m_1+m_2$ (for $m_1 \leq m_2 \leq m_3$). Therefore, in the case of three internal particles, two different analytical distributions arise depending on which of these conditions is satisfied. Taking this into account, the distribution $\rho^{\{ N=3 \}}(X)$ can be written as:
\begin{widetext}
 \[\rho^{\{ N=3 \}}(X)=Const \]
\begin{equation}\label{eq3}
\left\{\begin{array}{l}
I: \; \frac{(M X+(m_1+m_2+m_3)(X+R))^2}{2 m_1 m_2} \qquad X \in [\frac{-m_1-m_2-m_3}{M+m_1+m_2+m_3}R, \frac{m_1-m_2-m_3}{M+m_1+m_2+m_3}R]\\
II: \; 2R\frac{(M+m_1) X+(m_2+m_3)(X+R)}{m_2}  \qquad X \in [\frac{m_1-m_2-m_3}{M+m_1+m_2+m_3}R, \frac{-m_1+m_2-m_3}{M+m_1+m_2+m_3}R] \\
III: \; 4R^2-\frac{((m_1+m_2-m_3)R-(M+m_1+m_2+m_3)X)^2}{2 m_1 m_2}  \qquad  \left\{\begin{array}{l} 
 m_3 < m_1+m_2:\\
X \in [\frac{-m_1+m_2-m_3}{M+m_1+m_2+m_3}R, \frac{-m_1-m_2+m_3}{M+m_1+m_2+m_3}R ] \\
\\
m_3 > m_1+m_2:\\
X \in [\frac{-m_1+m_2-m_3}{M+m_1+m_2+m_3}R, \frac{m_1+m_2-m_3}{M+m_1+m_2+m_3}R ] \\
\end{array}\right.\\
IV: \left\{ \begin{array}{l}  m_3 < m_1+m_2:\\ 4R^2-\frac{(m_1+m_2-m_3)^2 R^2+(M+m_1+m_2+m_3)^2 X^2}{m_1 m_2}  \qquad X \in [\frac{-m_1-m_2+m_3}{M+m_1+m_2+m_3}R, \frac{m_1+m_2-m_3}{M+m_1+m_2+m_3}R ]  \\
m_3 > m_1+m_2:\\
4R^2 \qquad \qquad\qquad\qquad\qquad\qquad X \in [\frac{m_1+m_2-m_3}{M+m_1+m_2+m_3}R, \frac{-m_1-m_2+m_3}{M+m_1+m_2+m_3}R ]
\end{array}\right.\\
V: \; 4R^2-\frac{((m_1+m_2-m_3)R+(M+m_1+m_2+m_3)X)^2}{2 m_1 m_2} \qquad \left\{ \begin{array}{l}
m_3 < m_1+m_2:\\
X \in [\frac{m_1+m_2-m_3}{M+m_1+m_2+m_3}R, \frac{m_1-m_2+m_3}{M+m_1+m_2+m_3}R ] \\
\\
m_3 > m_1+m_2:\\
X \in [\frac{-m_1-m_2+m_3}{M+m_1+m_2+m_3}R, \frac{m_1-m_2+m_3}{M+m_1+m_2+m_3}R ] \end{array}{l}
\right.\\
VI: \; -2R\frac{(M+m_1) X+(m_2+m_3)(X-R)}{m_2} \qquad X \in [\frac{m_1-m_2+m_3}{M+m_1+m_2+m_3}R, \frac{-m_1+m_2+m_3}{M+m_1+m_2+m_3}R ] \\

VII: \; \frac{(M X+(m_1+m_2+m_3)(X-R))^2}{2 m_1 m_2} \qquad X \in [\frac{-m_1+m_2+m_3}{M+m_1+m_2+m_3}R, \frac{m_1+m_2+m_3}{M+m_1+m_2+m_3}R ] \\
\end{array}\right. 
\end{equation}
\end{widetext}
The normalization constant for both distributions is identical and is equal to:

\begin{equation}\label{eq4}
Const=\frac{M+m_1+m_2+m_3}{8 m_3 R^3}
\end{equation}

The distribution given by Eq. \ref{eq3} consists of smoothly joined segments that are polynomials of degree not higher than two. The formulas for the boundaries of these segments differ only in the signs of the masses appearing in the numerator of the expression of the form $\pm m_1 \pm m_2 \pm m_3$. In total $2^3$ such combinations are possible, so distribution with $N=3$ will give  $2^3-1$ segments. . In the more general case of an arbitrary number of internal particles, the boundaries of the segments of the distribution $\rho^{\{ N \}}(X)$ are determined by the conditions that the vertices of a hypercube with coordinates of the form $q_i=X \pm R$ lie on surfaces of the form  $\frac{M X+(X \pm R)\sum_{j=i+1}^{N} m_j +\sum_{j=1}^{i-1} m_j X_j}{m_i}$, where $X_j$ are $X_j=X \pm R$. It is easy to see that the resulting linear equation has solutions of the form $X_{lim}=R\frac{\sum_{i=1}^{N} \pm m_i}{M+\sum_{i=1}^{N} m_i}$. Therefore, for an arbitrary number of particles the distribution  $\rho^{\{ N \}}(X)$ consists of  $2^N-1$ segments, each being a polynomial of degree not higher than $N-1$. The specific form of the distribution depends on the ratios between the particle masses, in the same manner as in the case of  $N=3$.
\begin{figure}
 \centering
 \includegraphics[width=6 cm]{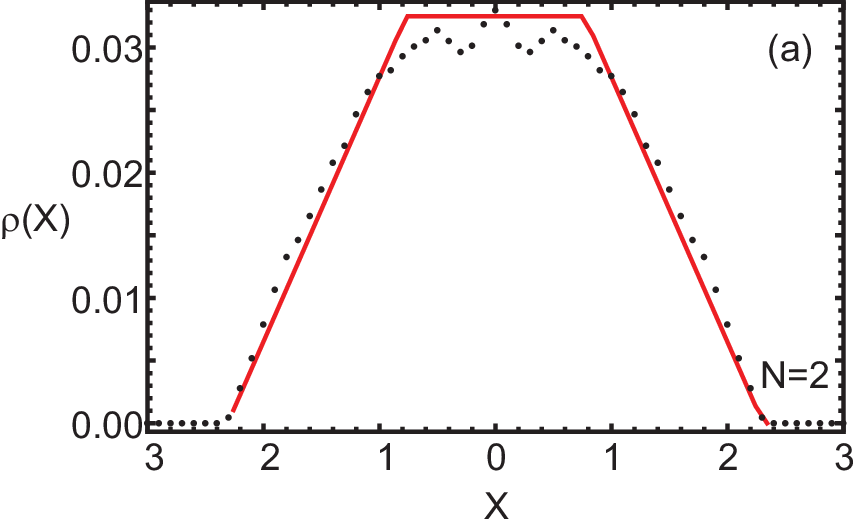}
 \includegraphics[width=6 cm]{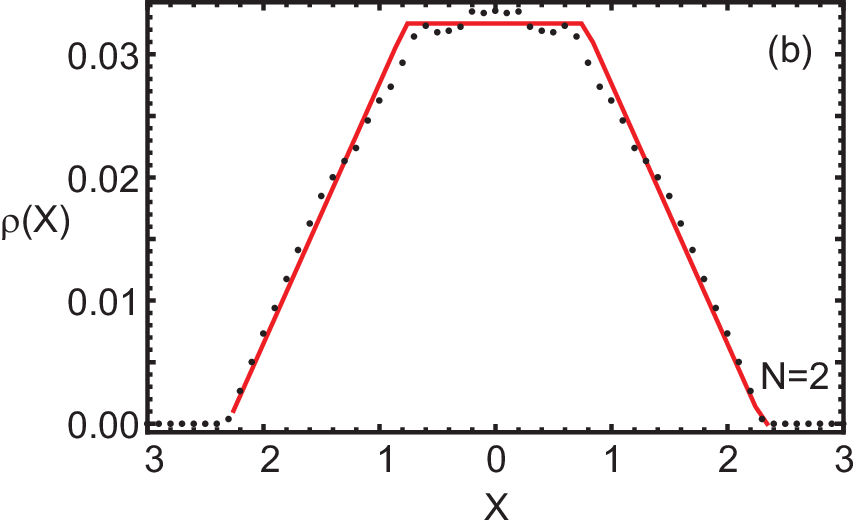}\\
 \includegraphics[width=6 cm]{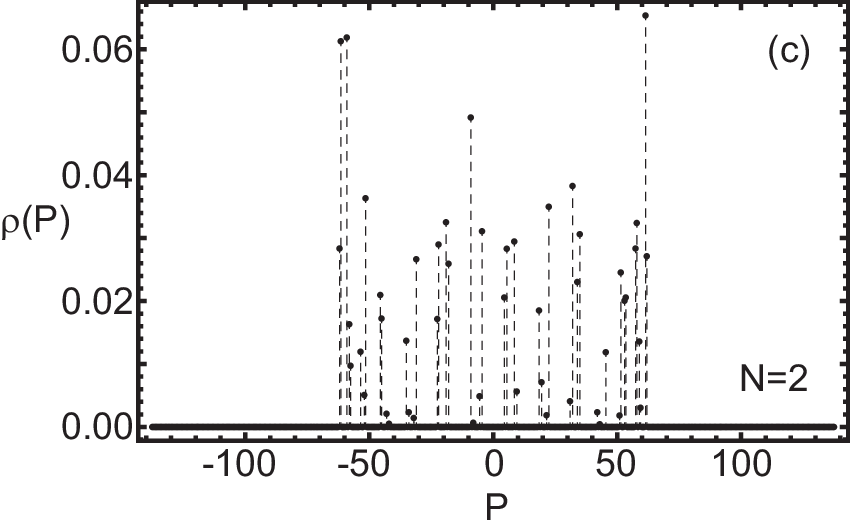}
 \includegraphics[width=6 cm]{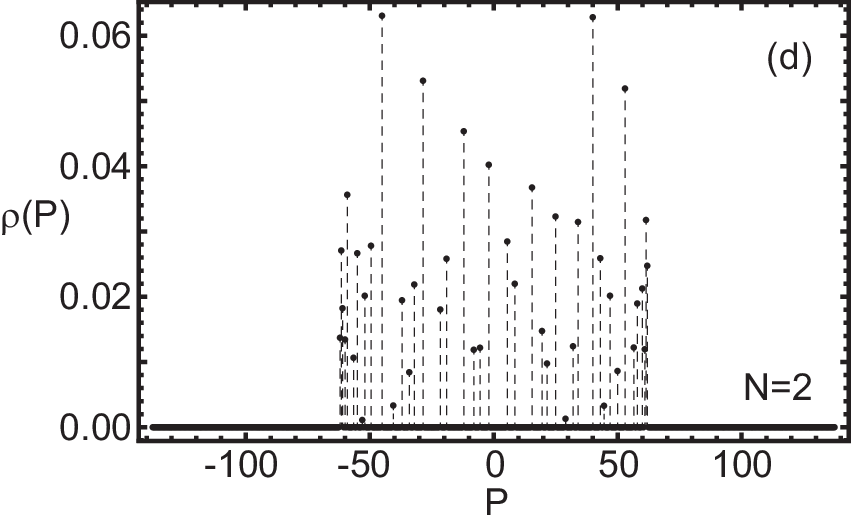}\\
 \caption{Typical distributions of the shell coordinate  (a,b) and the shell momentum (c,d) in the case of two non-colliding internal particles undergoing one-dimensional motion. The system parameters $M=3 \pi$, $m_1=1$, $m_2=\sqrt{2}$, $R=10$, as well as the values of the integrals of motion $E_{tot}=1000$, $P_{tot}=0$, $X_{tot}=0$, are identical for all plots. The lack of reproducibility and the deviation from the theoretical distribution indicate the existence of an additional integral of motion.}
 \label{ODO-fig3}
\end{figure}

\begin{figure}
 \centering
 \includegraphics[width=6 cm]{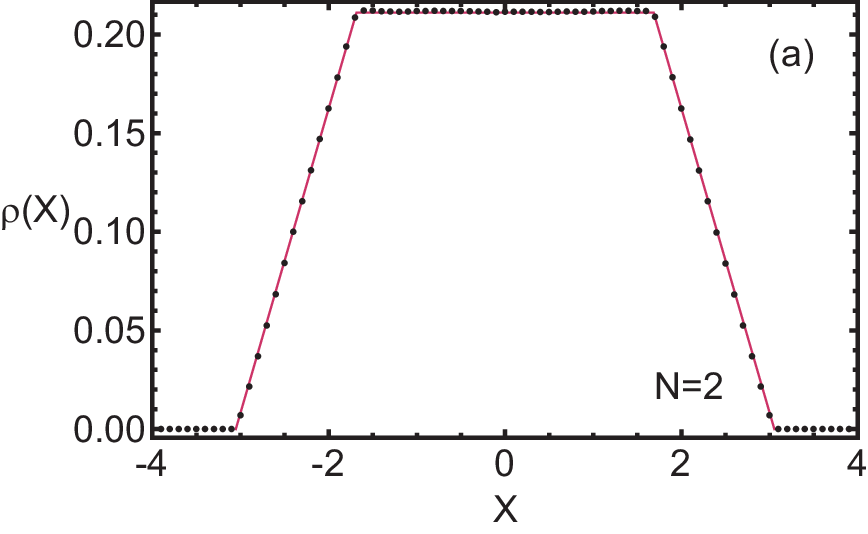}
 \includegraphics[width=6 cm]{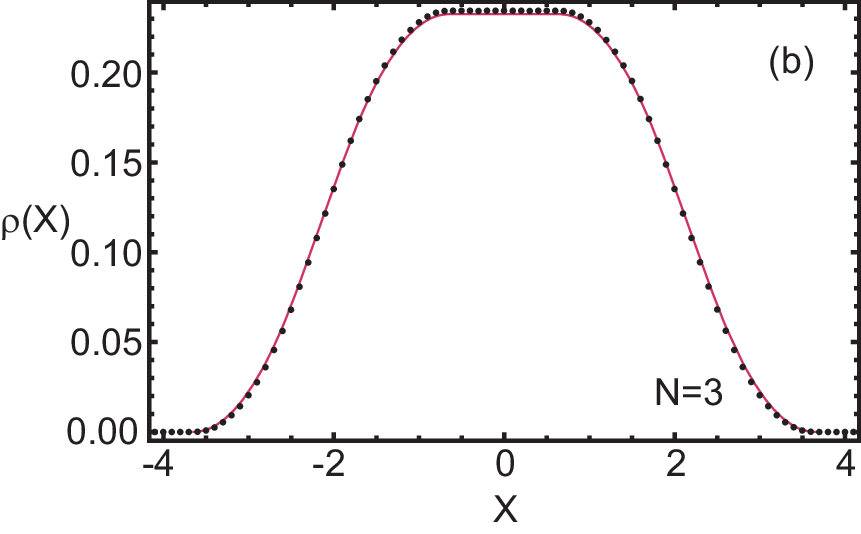}\\
 \includegraphics[width=6 cm]{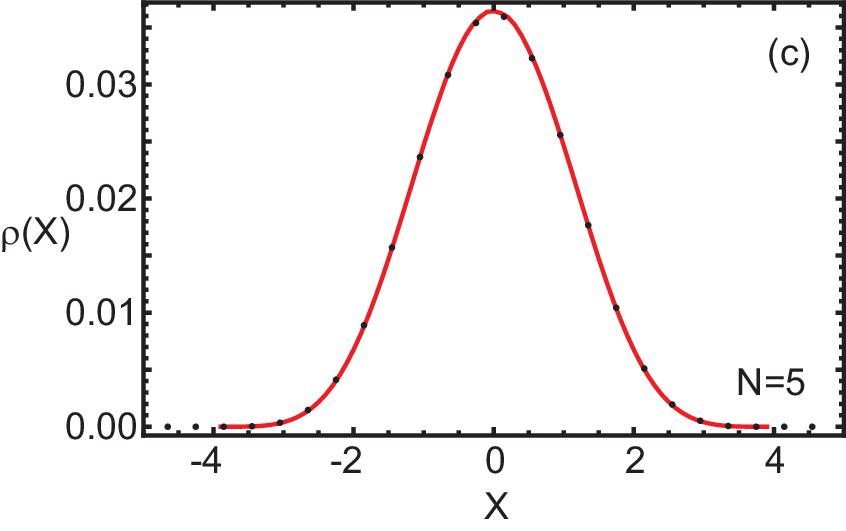}
 \includegraphics[width=6 cm]{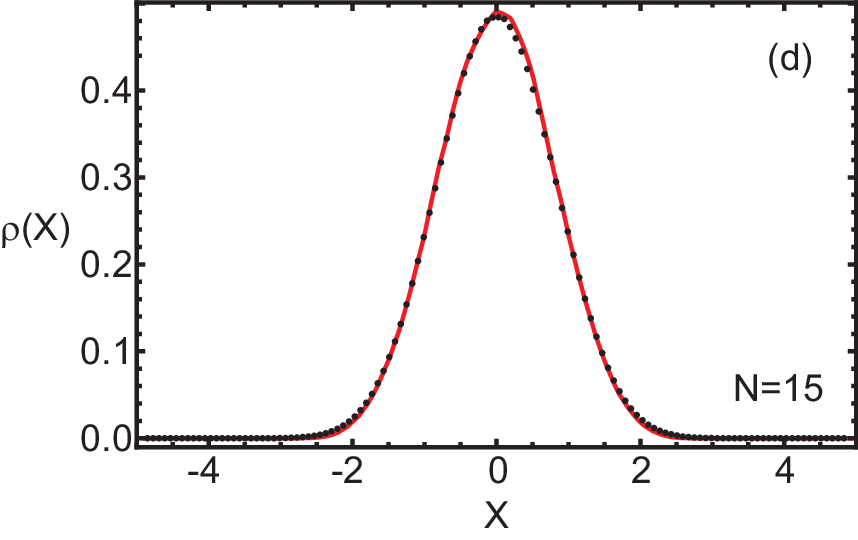}\\
  \caption{Distributions of the shell coordinate. The black dots represent the results of numerical simulations, while the red solid line corresponds to the distribution given by Eq. \ref{eq1}. The case of  $N=2$ and $N=3$ colliding internal particles  (a,b); (c,d) The case of  $N=5$ and $N=15$ non-colliding particles. The system parameters  $M=3 \pi$, $R=10$ and the values of the integrals of motion  $E_{tot}=1000$, $P_{tot}=0$, $X_{tot}=0$ were the same in all cases. Particle masses  (b) $m_1=1$, $m_2=\sqrt{2}$, $m_3=2\sqrt{3}$, (c) $m_1=1$, $m_2=\sqrt{2}$, $m_3=\sqrt{3}$, $m_4=\sqrt{5}/2$, $m_5=\sqrt{7}/3$ and (d) random within the range $m_i \in (0,1)$.}
 \label{ODO-fig4}
\end{figure}
Let us now compare the distributions obtained from Eqs.\ref{eq1} and \ref{eq3} with the results of numerical simulations. First, consider the case of a shell with two internal particles, shown in Fig.\ref{ODO-fig3}(a,b) and Fig.\ref{ODO-fig4}(a). It is evident that for non-colliding particles the experimental distributions differ significantly from the distribution given by Eq. \ref{eq1}. Moreover, for two internal particles these distributions depend on the choice of initial conditions. That is, even if the initial energy, momentum, and center-of-mass position are fixed, the resulting distributions change when the initial configuration is varied. Different initial conditions lead to different distributions, two examples of which are shown in Fig. \ref{ODO-fig3}(a,b). This clearly indicates that the dynamical equations are integrable and that there exists an additional integral of motion. As long as the value of this integral is not controlled and remains effectively random, the distributions corresponding to different initial conditions will differ. The distributions of the shell momentum for two internal particles are shown in Fig.\ref{ODO-fig3}(c,d). These distributions are also not reproducible and, moreover, are discrete. This discreteness arises because the three momenta of the system are related by three equations: conservation of energy, conservation of momentum, and one additional relation. The system of these equations has solutions consisting only of a finite set of values. Consequently, only discrete values of the shell momentum are possible, and the corresponding distribution becomes discrete.

In the case of two colliding particles, the distribution of the shell coordinate is shown in Fig. \ref{ODO-fig4}(a). In this situation the distribution becomes well defined and coincides with the theoretical prediction. As expected, collisions increase the degree of chaoticity in the motion and eliminate the additional integral of motion.

Although for  $N=2$ the system behavior is clearly influenced by the additional integral of motion, this effect disappears completely for a larger number of particles. Numerical simulations show that already for $N>2$ the distribution of the shell coordinate for non-colliding internal particles becomes uniquely defined and coincides with the theoretical prediction. A comparison of the numerically obtained distributions with Eq.\ref{eq3} is shown in Fig. \ref{ODO-fig2}(c,d). For colliding particles, the comparison is presented in Fig.\ref{ODO-fig4}(b). The simulation results agree completely with Eq. \ref{eq3}. This is also clearly demonstrated by the presence of a flat region of the distribution for those values of $X$, for which the integration domain is the entire square of side $2R$. Distributions for more than three particles were compared with the results of numerical integration of Eq. \ref{eq1} and are shown in Fig.\ref{ODO-fig4}(c,d). Thus, for  $N>2$ the numerical experiment confirms the assumptions made and demonstrates that the distribution of the shell coordinate is independent of the type of motion of the internal particles.

Knowing the exact form of the distribution $\rho^{\{ N=3 \}}(X)$, one can determine the root-mean-square deviation of the shell from the origin $<X^2>=\int_{X_{min}}^{X_{max}} X^2 \rho^{\{ N=3 \}}(X) dX$. For both cases of the distribution this value turns out to be the same:

\begin{equation}\label{eq5}
<X^2_{N=3}>=\frac{m_1^2+m_2^2+m_3^2}{(M+m_1+m_2+m_3)^2} \frac{R^2}{3}
\end{equation}

Despite the rather complicated structure of the shell coordinate distribution, the root-mean-square deviation has a simple form. This suggests that Eq. \ref{eq5} can be generalized to the case of an arbitrary number of particles as:

\begin{equation}\label{eq6}
<X^2_{N}>=\frac{\sum_{i=1}^{N} m_i^2 }{(M+m_{tot})^2} \frac{R^2}{3}
\end{equation}

\begin{figure}
 \centering
 \includegraphics[width=6 cm]{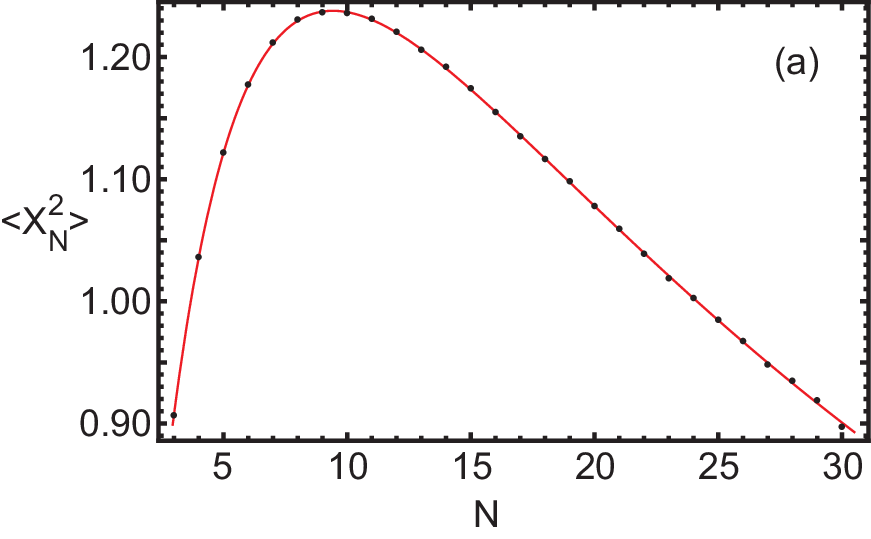}
 \includegraphics[width=6.3 cm]{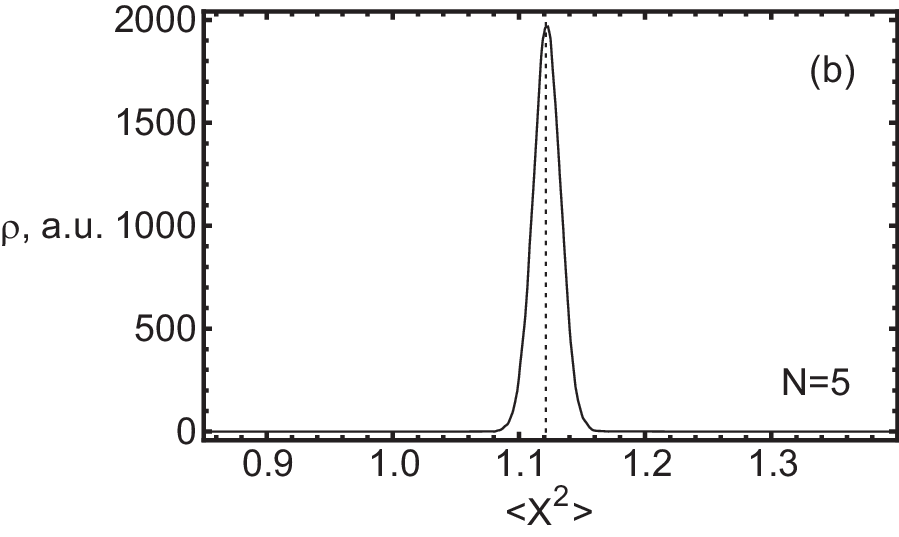}\\
  \caption{Standard deviation of the shell for different values of the number of non-colliding internal particles. The particle masses were chosen randomly, subject to the conditions  $m_{tot}=N$ and $\sum_{i=1}^{N} m_i^2=1.4 N$. (a) Dependence of the standard deviation on the number of particles. The dots represent simulation results, while the solid curve corresponds to Eq. (\ref{eq6}). (b) Distribution of the values of the standard deviation of the shell obtained from repeated simulations, each performed with different randomly chosen particle masses. It is evident that different mass configurations correspond to a single value of the standard deviation of the shell, around which the simulation results are clustered.}
 \label{ODO-fig5}
\end{figure}

To verify (\ref{eq6}) using simulation data, we begin with the case of non-colliding particles. In this case, the masses of the internal particles were generated randomly, but in such a way that the sums of the masses and the sums of their squares depended on the number of particles in a specific manner, namely $m_{tot}=N$ and $\sum_{i=1}^{N} m_i^2=1.4 N$. Under these conditions, Eq.(\ref{eq6}) ) acquires a definite dependence on $N$, these results can be compared with Eq.(\ref{eq6}). Figure \ref{ODO-fig5} (a) shows a comparison of the simulation data with Eq. (\ref{eq6}). A good agreement with the analytical relation  (\ref{eq6}) is observed.  Figure \ref{ODO-fig5}(b) presents fluctuations of the standard value at fixed  $N=5$ caused by the random selection of particle masses. 

\begin{figure}
	\centering
		\includegraphics[height=5 cm]{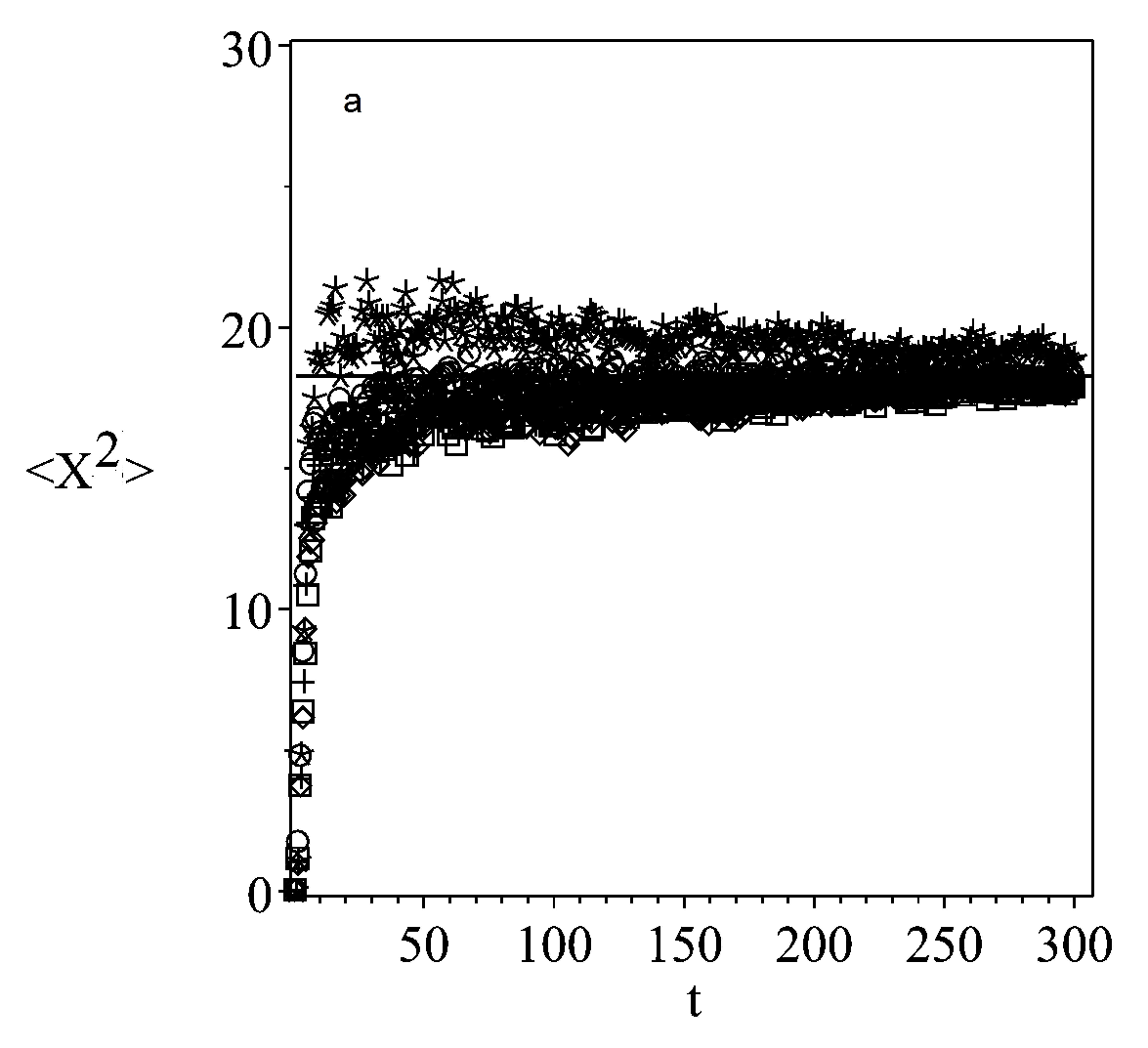}
		\includegraphics[height=5 cm]{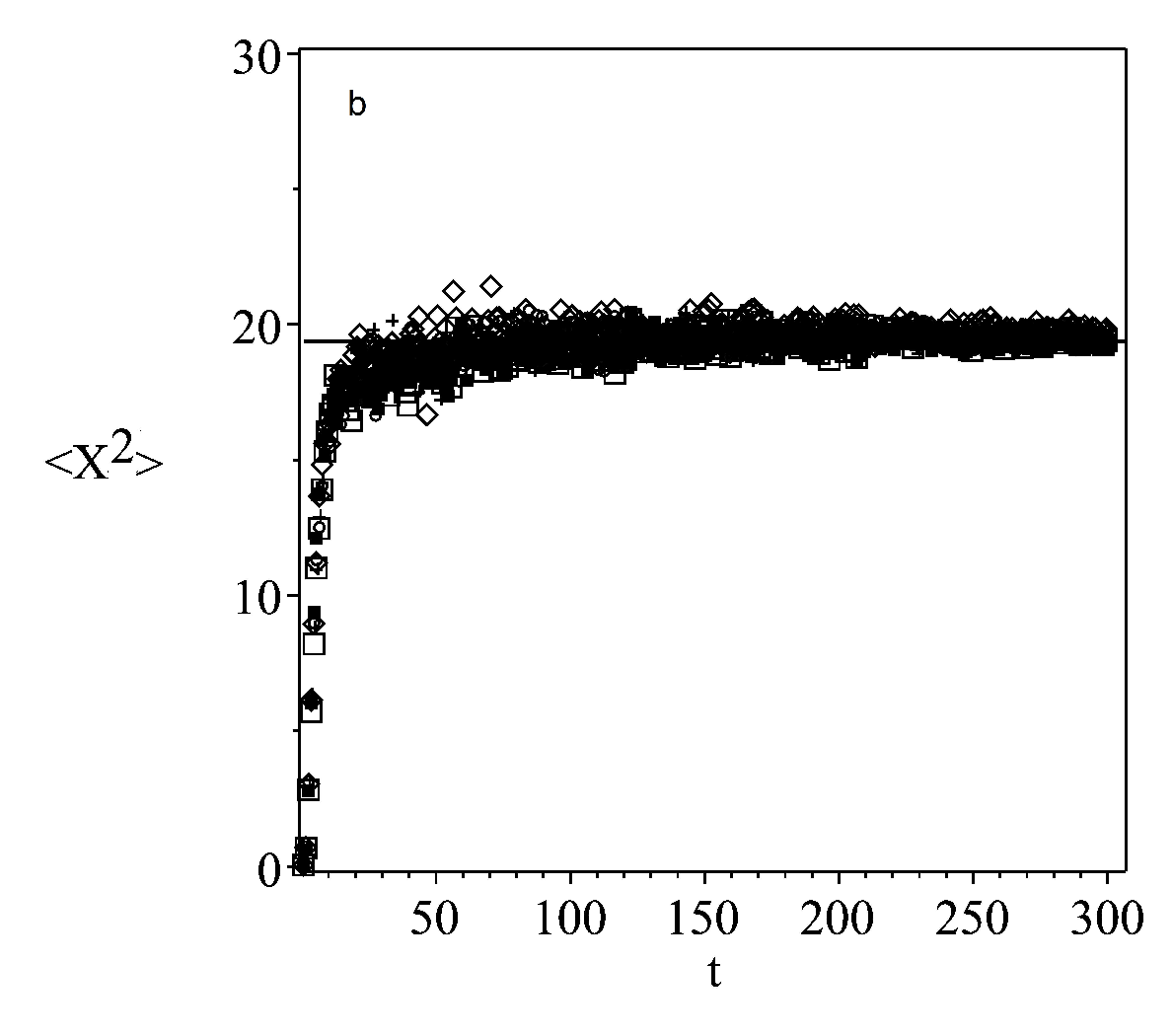}
	\caption{Convergence of the standard deviation to the value determined by relation (\ref{eq6}) or cases $N=5$(a) and $N=4$(b). The straight line indicates the value given by Eq.(\ref{eq6}), while different symbols represent the behavior for five different sets of masses. The sets of internal particle masses differ in individual values but have identical sums of masses and sums of squared masses for  $N=5$ $m_{tot}=28$ and $\sum_{i=1}^{N} m_i^2=198$ and $N=4$ $m_{tot}=28$ and $\sum_{i=1}^{N} m_i^2=210$. The time evolution of $\left\langle X^2 \right\rangle$ demonstrates convergence to the value given by  (\ref{eq6}). }
	\label{fig}
\end{figure}

The behavior of the standard deviation given by Eq.(\ref{eq6}) is also well confirmed by simulation data in the case of one-dimensional motion of colliding internal particles. Figure \ref{fig} shows the time evolution of the standard deviation of the shell with  $N=5$(a) and $N=4$(b) internal particles having different initial masses. Figure \ref{fig} presents the behavior for five different sets of particle masses, all having identical values of the sums of masses and sums of squared masses. Consequently, the value of the standard deviation given by Eq. (\ref{eq6}) is the same for all these cases. The approach to this value is illustrated in Fig. \ref{fig} For each mass set, 1000 simulations with different initial conditions were performed, and averaging over realizations was carried out. The resulting time dependence of the standard deviation is shown in Fig.\ref{fig} for the five mass sets. A relatively rapid relaxation toward the value determined by relation  (12)(Fig.\ref{fig}) is observed, along with small fluctuations.  

\begin{figure}
	\centering
		\includegraphics[width=5 cm]{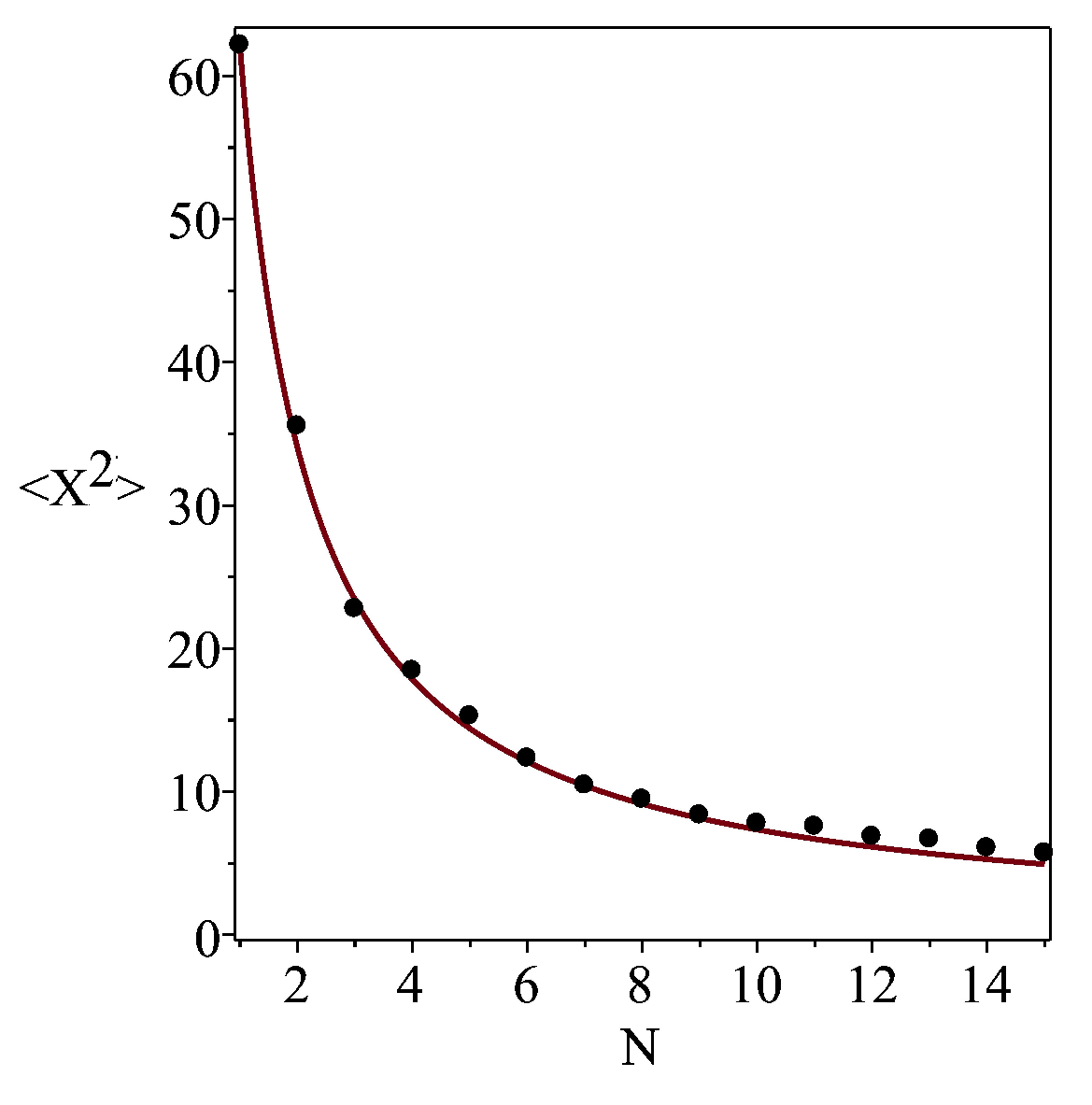}
	\caption{The standard deviation of the shell as a function of the number of internal particles is shown. The results of numerical simulations are represented by black circles, while the solid curve corresponds to the dependence given by Eq. \ref{eq7}. The simulation parameters are  $m=5$, $M=0.5$, $R=15$; collisions between particles were taken into account in the one-dimensional case. }
	\label{fig6n}
\end{figure}

Let us note another consequence of Eq.(\ref{eq6}). If the internal particle masses are chosen to be equal, a simple dependence of the root-mean-square deviation on the number of internal particles $N$ arises:
\begin{equation}\label{eq7}
<X^2_{N}>=\frac{{N} m^2 }{(M+ N m)^2} \frac{R^2}{3}
\end{equation}
Where $m$ is the mass of an internal particle. Thus, a specific dependence  $<X^2_{N}>$ on the number of internal degrees of freedom emerges. The results of numerical experiments for varying numbers of internal particles are shown in Fig.\ref{fig6n}. A good agreement between the simulation data and the dependence given by Eq. (\ref{eq7})is clearly observed.

Thus, for $N>2$ the distribution of the shell coordinate is universal and can easily be obtained in the form of a definite integral. An explicit analytical expression for this distribution can also be derived, although it is rather cumbersome, consisting of a large number of branches whose form depends on the ratios of the masses of the internal particles. Nevertheless, these complicated distributions lead to a simple law for the behavior of the root-mean-square deviation of the shell. It is important to emphasize that the chaotic behavior of the shell is determined by the universal relation (\ref{eq6}), which has a very general character. In a certain sense, it describes the chaotic motion of a "Brownian" particle in the absence of a surrounding medium.

\section{Distribution of the shell momentum}

Let us now consider the distribution of the shell momentum  $\rho(P)$ in the case where the internal particles undergo collisions with one another. In this case additional integrals of motion are absent; therefore, in order to compute this distribution, it is necessary to integrate the general probability density given by Eq. \ref{ODO-1-6} over all coordinates and all momentum components of the internal particles. Integration over the coordinates only changes the normalization constant of the distribution, since neither the coordinates appear in the integrand of Eq. \ref{ODO-1-6}, nor do the momenta enter the limits of integration over coordinates. The remaining integral over the particle momentum components takes the form of:
\begin{widetext}
\begin{equation}\label{ODO-eq10}
\rho(P) = \int\limits_{p_{x i}=p^{\{lim 1\}}_{x i}}^{p^{\{lim 2\}}_{x i}}\int\limits_{p_{y j}=-p^{\{lim\}}_{y j}}^{p^{\{lim\}}_{y j}} \frac{const \;\; dp_{x 1}..dp_{x N-2} \; dp_{y 1}..dp_{y N}}{\sqrt{\left( 2 E_{tot}-\frac{P^2}{M}-\sum\limits_{i=1}^{N-2}\frac{p_{x i}^2}{m_i}-\sum\limits_{j=1}^{N}\frac{p_{y j}^2}{m_i} \right)(m_{N-1} +m_N )-(P+\sum\limits_{i=1}^{N-2} p_{x i})^2}}
\end{equation}
\end{widetext}
where $x$-momentum components of the two last particles  $p_{x N-1}$ and $p_{x N}$. are chosen as the dependent variables. The integration limits for these momentum components are:
\begin{widetext}
\begin{equation}\label{ODO-eq11}
\begin{array}{l}
p^{\{lim 1,2\}}_{x i} = \sqrt{m_i} \frac{-\sqrt{m_i} (P+\sum\limits_{k=1}^{i-1} p_{x k}) \pm \sqrt{\sum\limits_{k=i+1}^{N} m_k} \sqrt{(2 E_{tot}-\frac{P^2}{M}-\sum\limits_{k=1}^{i-1}\frac{p_{x k}^2}{m_k}-\sum\limits_{j=1}^{N}\frac{p_{y j}^2}{m_j})(\sum\limits_{k=i}^{N} m_k)-(P+ \sum\limits_{k=1}^{i-1} p_{x k})^2}}{\sum\limits_{k=i}^{N} m_k}\\
\end{array}
\end{equation}
\end{widetext}
These expressions are most conveniently obtained from the condition that the integration is performed over the entire region in which the expression under the square root in Eq. \ref{ODO-eq10} is non-negative.

Integration over the  $x$-components can be performed sequentially using the following mathematical relation:
\begin{widetext}
\begin{equation}\label{ODO-eq12}
\begin{array}{l}
\int\limits_{p_{x i}=p^{\{lim 1\}}_{x i}}^{p^{\{lim 2\}}_{x i}} \left( (2 E_{tot}-\frac{P^2}{M}-\sum\limits_{k=1}^{i}\frac{p_{x k}^2}{m_k}-\sum\limits_{j=1}^{N} \frac{p_{y j}^2}{m_i})(\sum\limits_{k=i+1}^{N} m_k )-(P+\sum\limits_{k=1}^{i} p_{x k})^2 \right)^{A-\frac{1}{2}} dp_{x i} = \\

= C \left( (2 E_{tot}-\frac{P^2}{M}-\sum\limits_{k=1}^{i-1}\frac{p_{x k}^2}{m_k}-\sum\limits_{j=1}^{N}\frac{p_{y j}^2}{m_i})(\sum\limits_{k=i}^{N} m_k )-(P+\sum\limits_{k=1}^{i-1} p_{x k})^2 \right)^A\\
\end{array}
\end{equation}
\end{widetext}
Thus, integration over a given momentum component within the specified limits preserves the structure of the integrand. The result of the integration is simply that the momentum component with respect to which the integration was performed is removed from the expression. . At the same time, the mass of the corresponding particle is added to the total mass, and the exponent of the entire expression increases by $1/2$. When integrating over the next $x$-component of the internal particles, the same procedure is repeated, and so on. As a result, after integrating over all momentum components, the distribution takes the form:
\begin{widetext}
\begin{equation}\label{ODO-eq13}
\rho(P)=const \int\limits_{p_{y j}=-p^{\{lim\}}_{y j}}^{p^{\{lim\}}_{y j}}  \left(2 E_{tot}-\frac{P^2}{M}-\frac{P^2}{m_{tot}}-\sum\limits_{j=1}^{N}\frac{p_{y j}^2}{m_i}\right)^{\frac{N}{2}-\frac{3}{2}} dp_{y 1}..dp_{y N}
\end{equation}
\end{widetext}
with integration limits:
\begin{figure}
 \centering
 \includegraphics[width=5.5 cm]{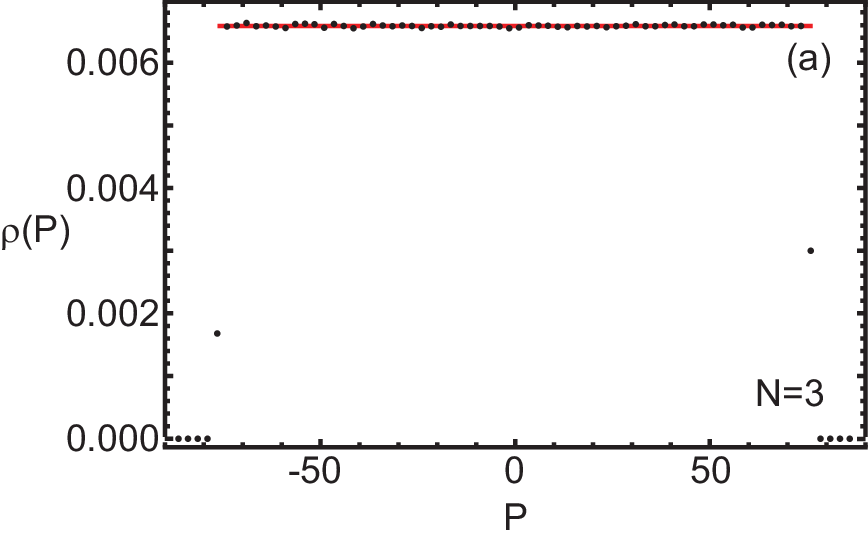}
 \includegraphics[width=5.5 cm]{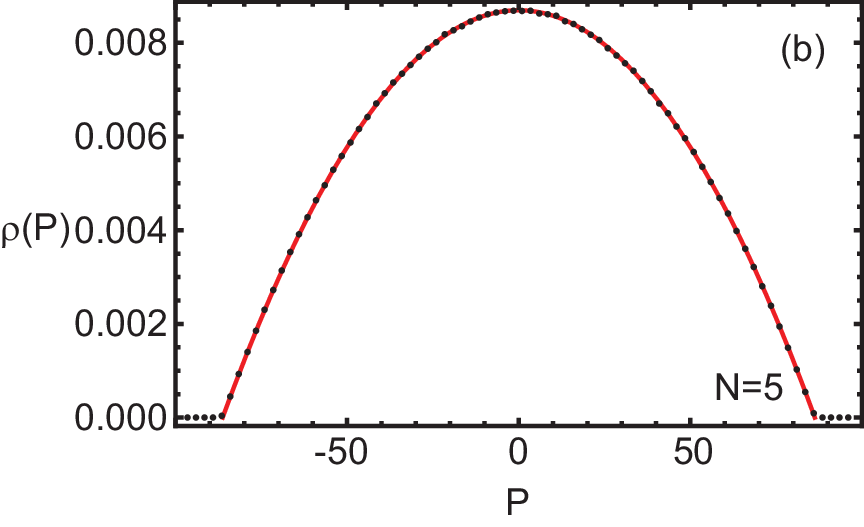}
 \includegraphics[width=5.5 cm]{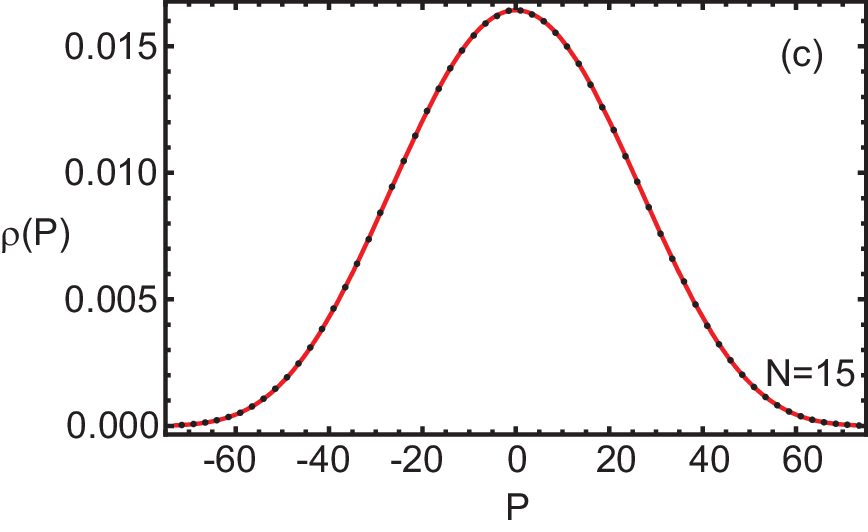}\\
 \includegraphics[width=5.5 cm]{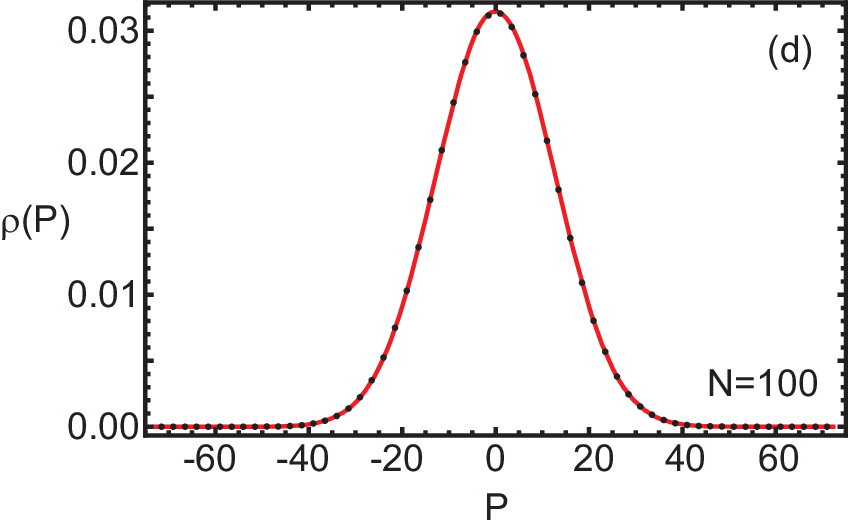}
 \includegraphics[width=5.5 cm]{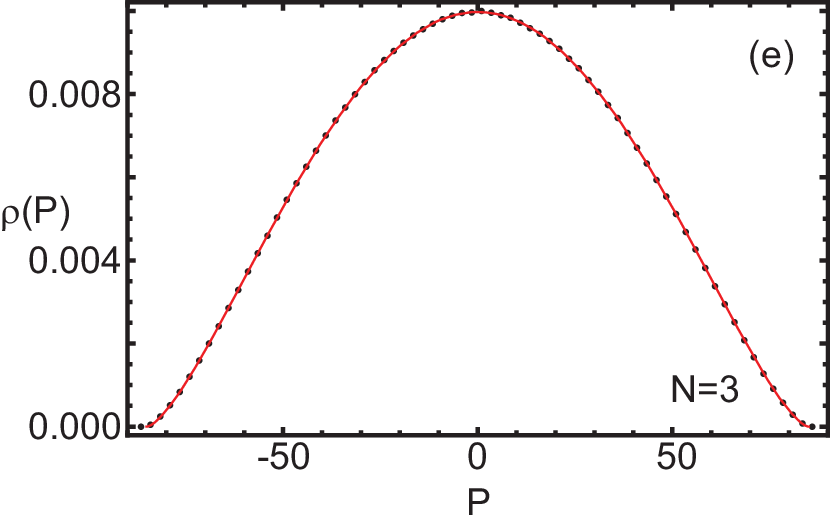}
 \includegraphics[width=5.5 cm]{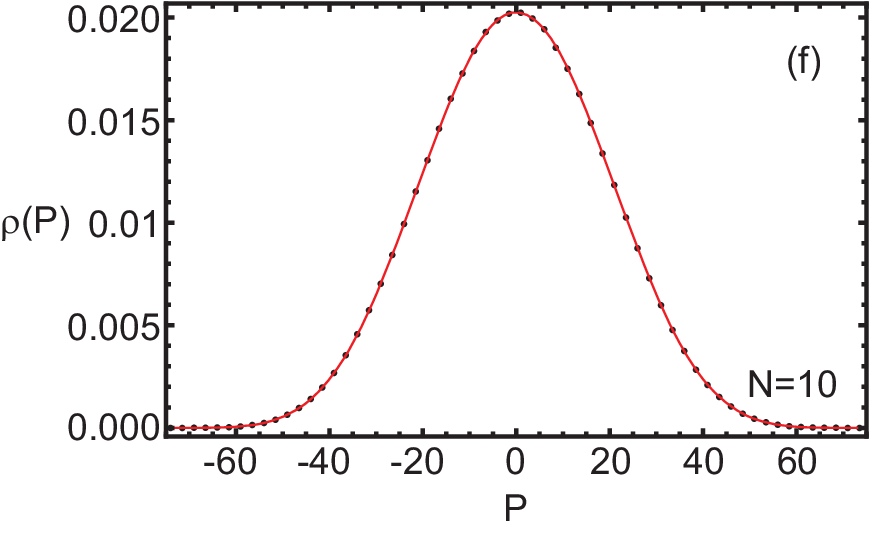}\\
  \caption{Distributions of the shell momentum in the case of more than two internal particles. The case of one-dimensional motion of the internal particles (a-d) the case of colliding particles (e,f) The shell parameters  $M=3 \pi$, $R=10$ as well as the values of the integrals of motion as well as the values of the integrals of motion  $E_{tot}=1000$, $P_{tot}=0$, $X_{tot}=0$ are identical in all cases.}
 \label{ODO-fig7}
\end{figure}
\begin{equation}\label{ODO-eq135}
p^{\{lim\}}_{y j} = \sqrt{m_j}\sqrt{2 E_{tot}-\frac{P^2}{M}-\frac{P^2}{m_{tot}}-\sum\limits_{k=1}^{j-1}\frac{p_{y k}^2}{m_k}}
\end{equation}

Here $\frac{P^2}{2 m_{tot}}$ is the minimal energy that must remain for the particles so that their $x$-components can compensate the shell momentum  $P$.

Further integration proceeds in the same manner as in the case where only the total energy of the system is conserved. Each integration step increases the exponent of the integrand by  $1/2$, without changing its structure. As a result, the distribution of the shell momentum can be obtained explicitly as:

\begin{equation}\label{ODO-eq14}
\begin{array}{l}
\rho(P)=C \left( E_{tot}  - \frac{(M+m_{tot}) P^2}{2 M m_{tot}} \right)^{N-\frac{3}{2}}\\
\\
C=\sqrt{\frac{M+m_{tot}}{2 \pi M m_{tot}}} \frac{\Gamma(N)}{E_{tot}^{N-1} \Gamma(N-\frac{1}{2})} \\
\end{array}
\end{equation}

The distribution lies within the interval $P \in [-\sqrt{\frac{2 E_{tot} M m_{tot} }{M+m_{tot}}},\sqrt{\frac{2 E_{tot} M m_{tot} }{M+m_{tot}}}]$. Qualitatively, this distribution has the same structure as the distributions of momentum components in systems where only the energy is conserved and no additional integrals of motion are present. The differences are that the shell mass is replaced by the reduced mass, and the exponent would be higher by $1/2$ in the case where only energy conservation is taken into account.

To obtain consistently the form of the shell momentum distribution for the case of one-dimensional motion of the internal particles, one must know the explicit form of all integrals of motion. However, assuming that no additional integrals exist, or that they do not influence the form of the distributions, the result is obtained in a manner completely analogous to the case of two-dimensional motion. The final distribution differs from that for two-dimensional particle motion only by replacing $N$ with $\frac{N}{2}$, since in the system with one-dimensional internal motion the number of degrees of freedom is smaller by $N$ and each integration over them increases the exponent of the distribution by  $\frac{1}{2}$. The general expression valid for both cases can be written as:

\begin{equation}\label{eq10}
\begin{array}{l}
\rho(P)=C \left( E_{tot} - \frac{(M+m_{tot}) P^2}{2 M m_{tot}} \right)^{\frac{D N-3}{2}}\\
\\
C=\sqrt{\frac{M+m_{tot}}{2 \pi M m_{tot}}} \frac{\Gamma(\frac{D N}{2})}{E_{tot}^{\frac{D N}{2}-1} \Gamma(\frac{D N-1}{2})} \\
\end{array}
\end{equation}
where $D$ - is the dimensionality of the motion of the internal particles. Thus, the form of the shell momentum distribution depends on the number of degrees of freedom of the system rather than on the number of internal particles.

Comparison of the distributions given by Eq. \ref{eq10} with the results of numerical simulations is shown in Fig.\ref{ODO-fig7}. It is seen that the shell momentum distributions for one-dimensional internal particle motion, shown in Fig. \ref{ODO-fig7}(a-d) coincide completely with Eq. \ref{eq10}. While for two internal particles the theoretical and experimental distributions do not coincide, already for three internal particles the theoretical distribution is fully confirmed. Moreover, for $N=3$ the shell momentum distribution becomes constant, meaning that the shell is equally likely to have any admissible momentum. It is interesting to note that for $N=2$ the distribution given by Eq.\ref{eq10} ) would qualitatively change its shape, diverging at the boundaries of the allowed momentum range. Comparison of Eq. \ref{eq10}with numerical simulations in the case of two-dimensional motion of internal particles is shown in Fig.\ref{ODO-fig7}(e,f). Again, the simulation results confirm Eq\ref{eq10}.

Knowing the analytical form of the momentum distribution, it is straightforward to obtain the distribution of the shell energy. In the general case it has the form:

\begin{equation}\label{eq11}
\begin{array}{l}
\rho(E)=\frac{C}{\sqrt{E}} \left( E_{tot}  - \frac{M+m_{tot}}{m_{tot}} E \right)^{\frac{D N-3}{2}}\\
\\
C=\sqrt{\frac{M+m_{tot}}{\pi m_{tot}}} \frac{\Gamma(\frac{D N}{2})}{E_{tot}^{\frac{D N}{2}-1} \Gamma(\frac{D N-1}{2})} \\
\end{array}
\end{equation}

This distribution lies within the $E \in [0,\frac{m_{tot}}{M+m_{tot}} E_{tot}]$ interval. From this distribution one can easily compute the average energy of the shell:

\begin{equation}\label{eq12}
<E>=\frac{m_{tot}}{D N (M+m_{tot})} E_{tot}
\end{equation}

As expected, the energy is not distributed uniformly among the degrees of freedom of the system. In the typical case  $D N>\frac{m_{tot}}{M}$ the average shell energy turns out to be lower than the equipartition value  $\frac{E_{tot}}{D N+1}$. As the shell mass increases, its average energy decreases, which appears to be characteristic of non-uniform distributions of kinetic energy. It is interesting to note that in the case of one-dimensional motion the equipartition value would occur only when all particle masses and the shell mass are equal. Only in this degenerate case does Eq. \ref{eq12} cease to apply.

\section{Discussion and Conclusions}

In this work, an exact analytical form of the coordinate and momentum distributions of a movable shell of a small system containing a finite number of internal particles has been obtained. The coordinate distribution was found to be largely independent of both the character and the dimensionality of the internal particle motion. This distribution is affected by the number of internal particles and their masses. It is also unaffected by integrals of motion that do not include the particle coordinates, such as the conservation of momentum. Despite this universality, the coordinate distribution of the shell has a rather complex structure. It consists of  $2^N-1$ branches, depends on all particle masses, and changes its analytical form depending on the ratios between the masses of the internal particles. A redistribution of mass among the internal particles may substantially modify the shape of the distribution. At the same time, the standard deviation of the shell is described by a simple universal formula and does not depend on the individual particle masses or on the total energy of the system.

The momentum distribution of the shell has a simple form and does not depend on the masses of the individual internal particles, but only on their total mass and on the mass of the shell itself. However, this distribution does depend on the number of degrees of freedom of the internal particles. Conservation of momentum, as in other analogous situations, leads to a modification of this distribution that results in a violation of the equipartition of energy among the degrees of freedom. The momentum distribution is described by a single function, whereas the coordinate distribution consists of several branches. It is interesting to note that in the case where both the energy and angular momentum $L_{tot}=0$, of a gas are conserved, the opposite situation occurs: the coordinate distribution is described by a single function, whereas the particle energy distribution consists of several branches.

Thus, the coordinate and momentum distributions of the shell are largely independent of one another. The same momentum distribution of the shell corresponds to an infinite number of coordinate distributions with different standard deviations. Therefore, the average energy of the shell is not related to its standard deviation, unless the shell coordinates explicitly enter the Hamiltonian of the system. The coordinate distribution of the system contains information about the number of particles and their masses, whereas the momentum distribution of the shell contains information about the number of internal degrees of freedom.

The obtained distribution functions make it possible to describe the chaotic motion of a structurally complex particle possessing a finite number of internal degrees of freedom. Universal relations describing the standard deviations of the shell have been derived. In a certain sense, this relation is analogous to the standard deviation of a Brownian particle. However, unlike Brownian motion, it is produced by impulses originating from the internal degrees of freedom rather than by collisions with molecules of an external medium. Consequently, this type of chaotic motion has a very general character and persists even in the absence of a surrounding medium. At the same time, it is naturally limited in magnitude. In addition, exact relations determining the distribution of energy among the degrees of freedom and between the internal particles and the shell have been obtained.

Acknowledgment
This work was supported by the National Research Foundation of
Ukraine, grant number 2025.07/0030 "Investigation of fundamental statistical manifestations of a finite number of internal degrees of freedom of nanoparticles".

\end{document}